\documentclass[twoside]{elsart}
\usepackage{epsfig}
\usepackage{floatflt}
\usepackage{amssymb,amsbsy,amsmath}
\newcommand{\xref}[1]{\protect\ref{#1}}

\newcommand{\figref}[1]{fig.~\protect\ref{#1}}
\newlength{\CaptionWidth}
\newcommand{\mycaption}[2]{%
\begin{center}\begin{minipage}{\CaptionWidth}%
\caption[]{#1}\label{#2}%
\end{minipage}\end{center}%
}%
\newcommand{\element}[2]{$^{#1}$#2}

\newcommand{\fm}{\mbox{fm}}
\newcommand{\MeV}{\mbox{MeV}}

\newcommand{\Operator}[1]{\raisebox{-1.1ex}{
$\!\!\stackrel{\displaystyle #1}{\sim}$}}

\def\half{\frac{1}{2}\;}
\def\bra#1{\langle \, {#1} \, | \;}
\def\ket#1{\; | \, {#1} \, \rangle}

\newcommand{\braket}[2]{\langle \, {#1} \, | \, {#2} \, \rangle}

\begin{document}
%
\typeout{   --- >>>   GSI-Preprint-97-18   <<<   ---   }
\typeout{   --- >>>   GSI-Preprint-97-18   <<<   ---   }
\typeout{   --- >>>   GSI-Preprint-97-18   <<<   ---   }
\begin{frontmatter}
\title{The nuclear liquid--gas phase transition\\ 
       within Fermionic Molecular Dynamics}
 
\author{J. Schnack\thanksref{JS}}
\author{ and H. Feldmeier\thanksref{HF}}
\address{Gesellschaft f\"ur Schwerionenforschung mbH, \\ 
         Postfach 110 552, D-64220 Darmstadt \&\\
         Technische Hochschule Darmstadt}

\thanks[JS]{email: j.schnack\char'100gsi.de,
            WWW:~http://www.gsi.de/$\sim$schnack}
\thanks[HF]{email: h.feldmeier\char'100gsi.de,
            WWW:~http://www.gsi.de/$\sim$feldm}

\begin{abstract}
The time evolution of excited nuclei, which are in equilibrium
with the surrounding vapour, is investigated. It is shown that
the finite nuclear systems undergo a first oder phase
transition.  The caloric curve is presented for excited
\element{16}{O}, \element{24}{Mg}, \element{27}{Al} and
\element{40}{Ca} and the critical temperature is estimated for
\element{16}{O}.\\[1mm]


\end{abstract}
\end{frontmatter}
\section{Introduction}
\label{Intro}

Mean--field models predict a first order phase transition
for nuclear matter with a critical temperature which depends on the
proton--neutron asymmetry \cite{JMZ84,GKM84,BLV84,SeW86,SCG89}.
A recent experimental attempt by the ALADIN group \cite{Poc95} to
deduce an equation of state, which relates the excitation energy
of a hot nucleus to its temperature, has stimulated both,
theoretical and experimental efforts in this field 
\cite{NHW95,MuS95,MoP96,PaN95,FBK96,Poc96,XLT96}.
In the experiment excited projectile spectators were
investigated in Au+Au collisions at a beam energy of
$E/A=600~\MeV$.  While the equation of state refers to a
stationary system where liquid and vapour (evaporated nucleons)
are in equilibrium, the experiment deals with an expanding
source.  This causes some uncertainties for the temperature,
which is deduced from isotope ratios, since the system cools
while it is expanding.

In molecular--dynamics calculations the finite system may be
excited without flow, but similar problems arise when the phase
transition sets in. Particles which escape from the nucleus cool
down the residue and thermal equilibrium cannot be maintained.
In order to avoid these difficulties, in the present simulations
the excited nuclear system is confined by a wide container
potential which is chosen to be a harmonic oscillator
potential. Its frequency $\omega$ serves as a thermodynamic
variable like the volume in the ideal gas case.  Due to the
containment evaporated nucleons cannot escape, but form a cloud
of equilibrated vapour around the excited nucleus.

Thermodynamic relations are obtained by coupling the nuclear
system to a reference system which serves as a thermometer.
Both, the time--evolution of the nuclear system and of the
thermometer are described by the Fermionic Molecular Dynamics
(FMD) model. Assuming thermal equilibrium in the sense of
ergodicity the temperature of the nuclear system is derived from
the time--averaged energy of the thermometer and related to the
excitation energy of the nucleus.

\section{Model and setup}
\subsection{The Fermionic Molecular Dynamics model}
\label{FMDModel}

The time evolution of the nuclear system is described within the
framework of Fermionic Molecular Dynamics (FMD) published in
detail in Ref. \cite{FBS95}.
The model describes the many--body system with a parameterized 
antisymmetric many--body state 
$\ket{Q(t)}$
composed of single--particle Gaussian wave packets
\begin{eqnarray}
\braket{\vec{x}}{q(t)} &=&
\exp\left\{-\frac{(\,\vec{x}-\vec{b}(t)\,)^2}{2\,a(t)}
    \right\}
\otimes\ket{m_s}\otimes\ket{m_t}
\ , \\
&&
\vec{b}(t) = \vec{r}(t) + i\, a(t)\, \vec{p}(t)
\nonumber
\ ,
\label{gaussian}
\end{eqnarray}
which are localized in phase space at $\vec{r}$ and $\vec{p}$
with a complex width $a$.  Spin and isospin are chosen to be
time--independent in theses calculations; they are represented
by their $z$--components $m_s$ and $m_t$, respectively.  Given
the Hamilton operator $\Operator{H}$ the equations of motion for
all parameters are derived from the time--dependent variational
principle (operators are underlined with a tilde)
\begin{eqnarray}
\delta \int_{t_1}^{t_2} \! \! dt \;
\bra{Q(t)}\; i \frac{d}{dt} - \Operator{H}\; \ket{Q(t)} \ &=&\ 0
\ .
\label{var}
\end{eqnarray}
In the present investigation the effective two--body
nucleon--nucleon interaction $\Operator{V}$ in the Hamilton
operator consists of a short--range repulsive and long--range
attractive central potential with spin and isospin admixtures
and includes the Coulomb potential \cite{Sch96}.  
The parameters of the interaction have been
adjusted to minimize deviations between calculated and measured
binding energies and charge radii for nuclei with mass numbers
$4\le A\le40$.

\subsection{The container}
\label{container}

Besides the kinetic energy $\Operator{T}$
and the nucleon--nucleon interaction $\Operator{V}$
the Hamilton operator $\Operator{H}$ includes an external field 
\begin{eqnarray}
\Operator{U}(\omega)
=
\half m \omega^2 \sum_{i=1}^A \vec{\Operator{x}}_i^2
\end{eqnarray}
which serves as a container.

The container is an important part of the setup because it keeps
the evaporated nucleons (vapour) in the vicinity of the
remaining liquid drop so that it equilibrates with the
surrounding vapour.  The vapour pressure is controlled by the
external parameter $\omega$, which appoints the accessible volume.

\subsection{The thermometer}
\label{thermometer}

The concept of determining the temperature is to bring a
reference system, for which thermodynamic relations between
temperature and measurable quantities are known, into thermal
equilibrium with the investigated system.  The weakly
interacting ideal gas, where the temperature is given by the
mean kinetic energy of the particles, may serve as an
example. The reference system is called a heat bath if its heat
capacity is much larger than that of the system and it is called
a thermometer if its heat capacity is much less.

As the nuclear system is quantal and strongly interacting its
temperature cannot be deduced from the momentum distribution or
the mean kinetic energy of the nucleons. Therefore, the concept
of an external thermometer which is coupled to the nuclear
system is used in the present investigation. The thermometer
consists of a quantum system of distinguishable particles moving in a
common harmonic oscillator potential different from the
container potential.

The time evolution of the whole system is described by the FMD
equations of motion. For this purpose the many--body trial
state is extended and contains now both, the nucleonic degrees
of freedom and the thermometer degrees of freedom
\begin{eqnarray}
\ket{Q}
=
\ket{Q_n} \otimes \ket{Q_{th}}
\ .
\end{eqnarray}
The total Hamilton operator including the thermometer is given
by
\begin{eqnarray}
\Operator{H}
=
  \Operator{H}_n
+ \Operator{H}_{th}
+ \Operator{H}_{n-th}
\ ,\quad
\Operator{H}_n
=
  \Operator{T}
+ \Operator{V}
+ \Operator{U}(\omega)
\end{eqnarray}
with the nuclear Hamilton operator $\Operator{H}_n$
and the thermometer Hamilton operator
\begin{eqnarray}
\Operator{H}_{th}
=
\sum_{i=1}^{N_{th}}
\left(
  \frac{\vec{\Operator{k}}^2(i)}{2\; m_{th}(i)}
+ \half m_{th}(i)\; \omega_{th}^2\; \vec{\Operator{x}}^2(i)
\right)
\ .
\end{eqnarray}
The coupling between nucleons and thermometer particles,
$\Operator{H}_{n-th}$, is assumed to be weak, repulsive and of
short range. It has to be as weak as possible in order not to
influence the nuclear system too much. On the other hand it has
to be strong enough to allow for reasonable equilibration
times. Our choice is to put more emphasis on small correlation
energies, smaller than the excitation energy, and to tolerate
long equilibration times.

The determination of the caloric curve is done in the following
way. The nucleus is excited by displacing all wave packets from
their ground--state positions randomly. Both, centre of mass
momentum and total angular momentum are kept fixed at zero. To
allow a first equilibration between the wave--packets of the
nucleus and those of the thermometer the system is evolved over
a long time (10000~\fm/c). After that a time--averaging of the
energy of the nucleonic system as well as of the thermometer is
performed over a time interval of 10000~\fm/c. During this time
interval the mean of the nucleonic excitation energy
\begin{eqnarray}
E^*
&=&
\frac{1}{N_{steps}} \sum_{i=1}^{N_{steps}}
\bra{Q_n(t_i)} \Operator{H}_n \ket{Q_n(t_i)}
- E_0(N,Z)
\end{eqnarray}
is evaluated.  $E_0(N,Z)$ denotes the FMD ground--state energy
of the isotope under consideration.  The time--averaged energy
of the thermometer $E_{th}$, which is calculated during the same
time interval, determines the temperature $T$ through the
relation for an ideal gas of distinguishable particles in a
common harmonic oscillator potential (Boltzmann statistics)
\begin{eqnarray}
T
&=&
\hbar\omega_{th}
\left[
\ln\left(
\frac{E_{th}/N_{th}+\frac{3}{2}\hbar\omega_{th}}
     {E_{th}/N_{th}-\frac{3}{2}\hbar\omega_{th}}
\right)
\right]^{-1}
\ .
\end{eqnarray}

\section{The caloric curve}
\label{Calorie}

The relation between the excitation energy and the
temperature is determined for the four nuclei 
\element{16}{O}, \element{24}{Mg}, \element{27}{Al} and
\element{40}{Ca} using the same container potential with
$\hbar\omega=1~\MeV$. In addition the dependence on $\omega$ is
investigated for \element{16}{O} leading to an estimate of the
critical temperature.

\begin{figure}[hhht]
\unitlength1mm
\begin{picture}(120,55)
\put(10, 1){\epsfig{file=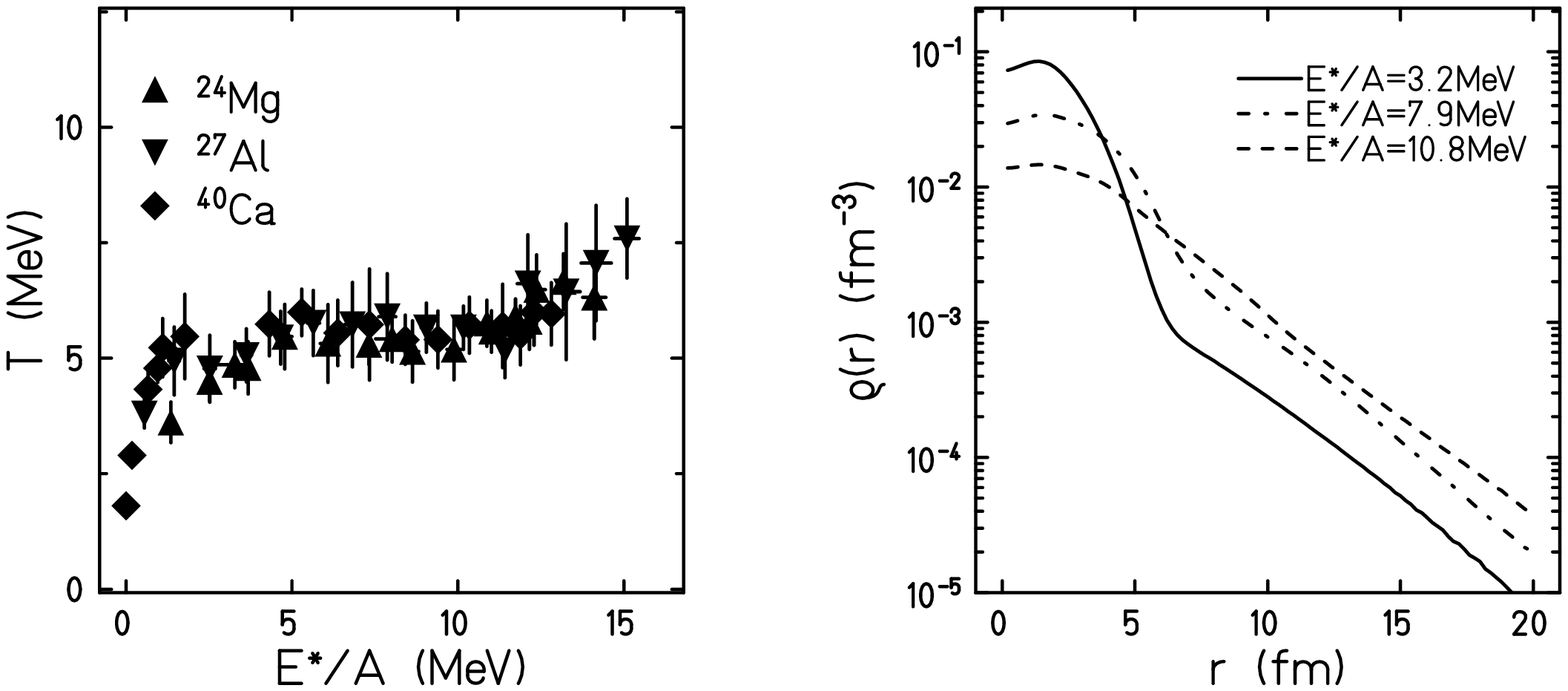,height=50mm}}
\end{picture}
\mycaption{L.h.s.: caloric curve of \element{24}{Mg}, 
\element{27}{Al} and 
\element{40}{Ca} at $\hbar\omega=1~\MeV$,
r.h.s.: time--averaged radial density distribution of\
\element{24}{Mg} at various excitation energies in the
coexistence region.}{CC-1}
\end{figure} 

The caloric curve shown in the graph on the left hand side of
fig.~\xref{CC-1} clearly exhibits three different
parts. Beginning at small excitation energies the temperature
rises steeply with increasing energy as expected for the shell
model. The nucleons remain bound in the excited nucleus which
behaves like a drop of liquid.  At an excitation energy of
$3~\MeV$ per nucleon the curve flattens and stays almost
constant up to about $11~\MeV$.  This plateau at
$T\approx5~\MeV$ has its origin in the coexistence of liquid and
vapour phases, the latter consisting of evaporated nucleons
which are in equilibrium with the residual liquid drop due to
the containment.  At the beginning of the plateau the system
contains only few evaporated nucleons, towards the end more and
also small intermittent condensed fragments which may amalgamate
or dissolve into vapour again.  Around $E^*/A\approx11~\MeV$ all
nucleons are unbound and the system has reached the vapour
phase. This is reflected by the rise of the caloric curve
beyond this point. The "error bars" denote the
r.m.s. fluctuations in temperature and excitation energy,
respectively, which arise from the energy exchange between
thermometer and nucleons. Only the total energy of the system
(nucleons \& thermometer) is conserved. The magnitude of the
temperature fluctuations is larger than those of $E^*/A$ because
the heat capacity of the thermometer is smaller than that of the
nucleus.

One has to keep in mind that the plateau, which due to finite
size effects is rounded, is not the result of a Maxwell
construction as in nuclear matter calculations.  In the
excitation energy range between 3 and $11~\MeV$ per particle an
increasing number of nucleons is found in the vapour phase
outside the liquid phase.  This can be seen in the density plot
on the right hand side of figure~\xref{CC-1}, where the
radial dependence of the time--averaged density of a system of
24 nucleons is shown at three excitation energies in the
coexistence region. For small excitations ($E^*/A=3.2~\MeV$) the
nucleus, which is bouncing around due to recoil, is surrounded
with very low density vapour (solid line).  The dashed--dotted line
($E^*/A=7.9~\MeV$) and the dashed line
($E^*/A=10.8~\MeV$) show that with increasing energy the vapour
contribution is growing and the amount of liquid decreasing.
However, in the high energy part of the plateau the averaged
one--body density displayed here does not represent the physical
situation adequately. The time--dependent many--body state shows
the formation and disintegration of several small drops. Above
$E^*/A\approx13~\MeV$ only vapour is observed.

\begin{floatingfigure}{60mm}
\epsfig{file=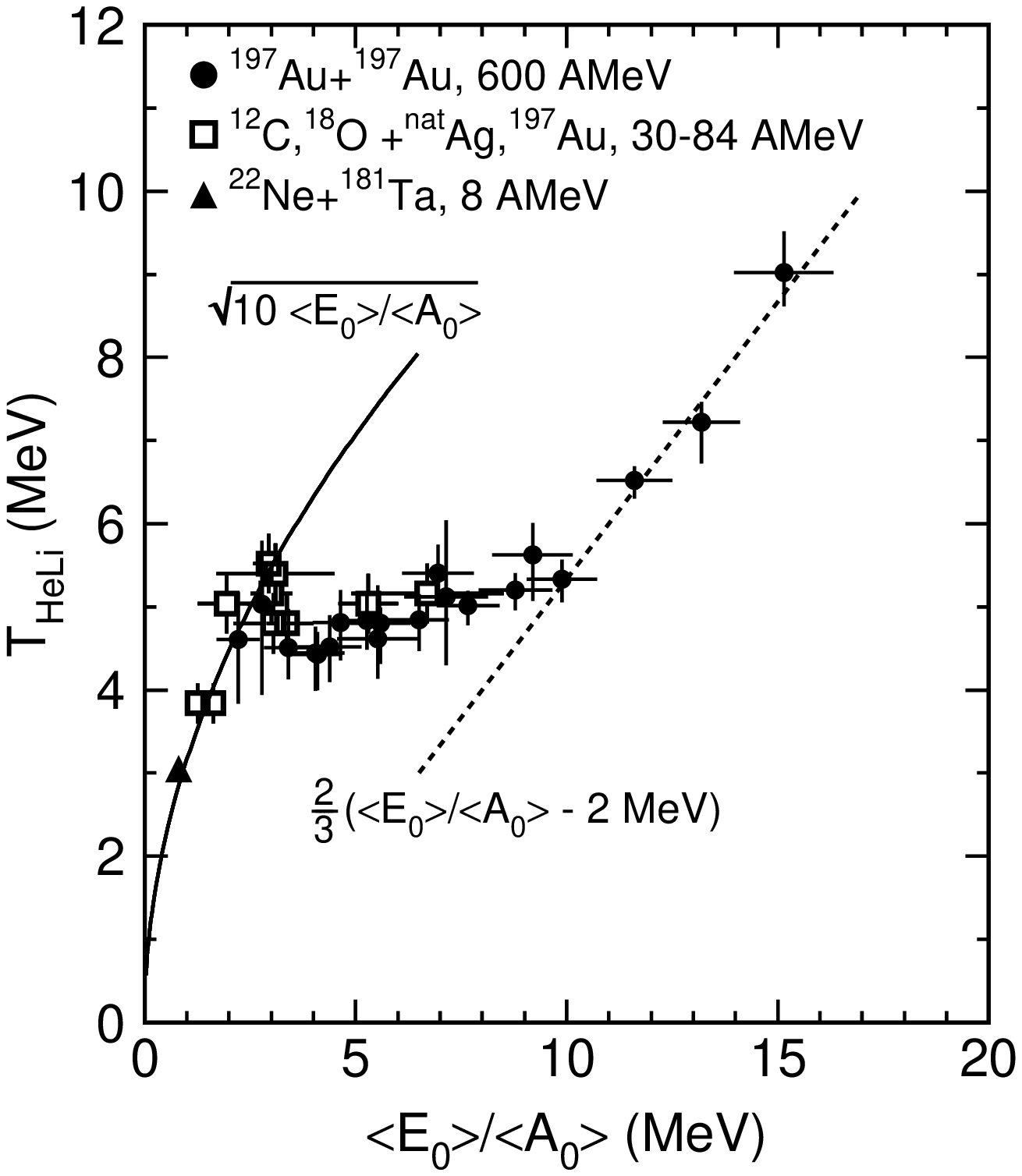,height=70mm}
\caption{Caloric Curve determined by the ALADIN group \cite{Poc95}.}
\label{CC-3}
\end{floatingfigure}
The caloric curve shown in \figref{CC-1} has a striking
similarity with the caloric curve determined by the ALADIN group
\cite{Poc95} which is displayed in \figref{CC-3}. The position
and the extension of the plateau agree with the FMD calculation
using a containing oscillator potential of
$\hbar\omega=1~\MeV$. Nevertheless, there are important
differences. The measurement addresses an expanding
non-equilibrium system, but the calculation deals with a
contained equilibrium system. In addition the used thermometers
differ; the experiment employs an isotope thermometer based on
chemical equilibrium and the calculation uses an ideal gas
thermometer.  One explanation why the thermodynamic description
of the experimental situation works and compares nicely to the
equilibrium result might be, that the excited spectator matter
equilibrates faster into the coexistence region \cite{FeS97a}
than it expands and cools. The assumption of such a transient
equilibrium situation \cite{PaN95,BDM85,Gro90} seems to work
rather well at least in the plateau region. Further
investigations on the FMD time evolution of excited nuclei
without container will focus on such assumptions.

The thermodynamic properties of the nucleonic system are
controlled by the external parameter $\omega$ similar to the
volume in the case of an ideal gas. An increasing $\omega$ leads
to smaller volume and to higher pressure in the container which
shifts the plateau of the caloric curve to higher temperatures
and decreases its extension, i.e. the latent heat. The critical
temperature $T_c$ is reached for the $\omega_c$ at which the
plateau vanishes.  Figure \xref{CC-2} displays this dependence
for \element{16}{O}. The caloric curve is evaluated for three
different oscillator frequencies of the container potential.  A
pronounced plateau is seen in the plot on the left hand side,
where the oscillator does not influence the self bound nucleus
very much. In the middle part the more narrow container
potential is already squeezing the ground state, its energy goes
up to $E/A\approx -5~\MeV$. The plateau is shifted to $T\approx
7~\MeV$ and the latent heat is decreased. On the right hand
side, for $\hbar\omega=18~\MeV$, the coexistence region has
almost vanished.
In addition one observes very large fluctuations of $T$ and $E$.

\begin{figure}[hhht]
\unitlength1mm
\begin{picture}(150,55)
\put(2,1){\epsfig{file=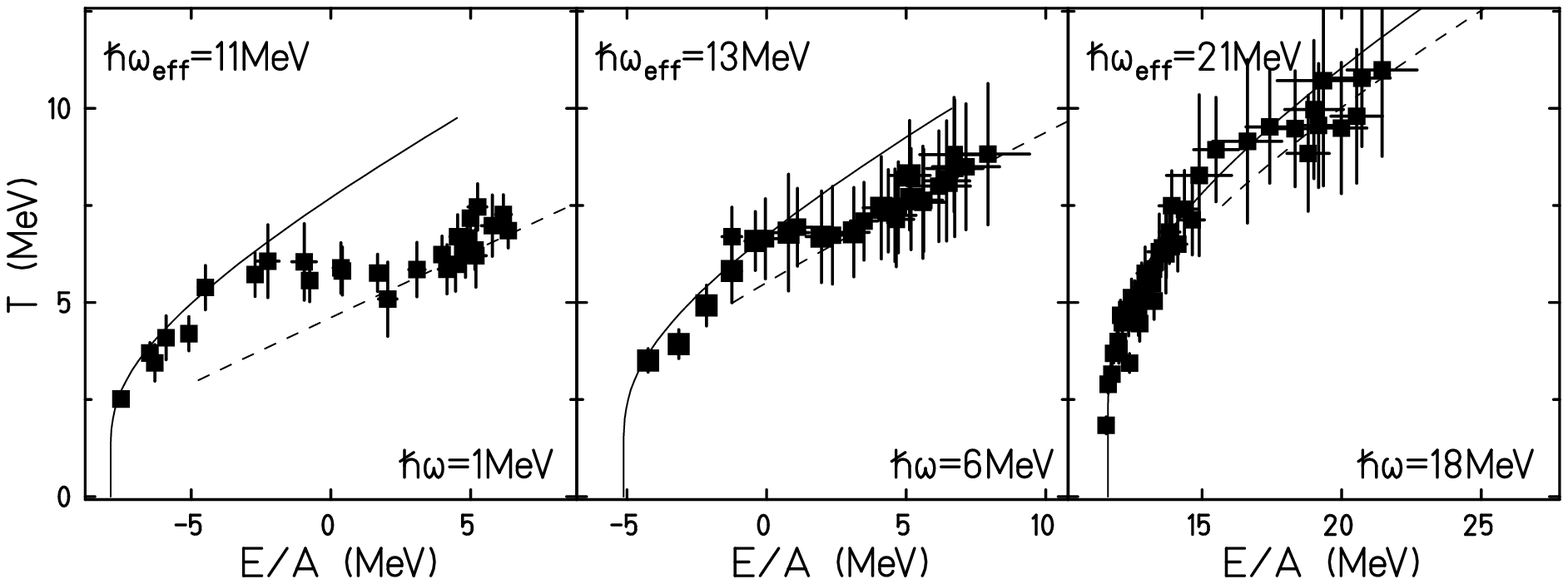,height=50mm}}
\end{picture}
\mycaption{Caloric curve of \element{16}{O} for the
frequencies $\hbar\omega=1, 6, 18~\MeV$ of the container
potential.  The solid lines show the low temperature behaviour
of an ideal gas of 16 fermions in a common harmonic oscillator
with level spacing $\hbar\omega_{\mbox{eff}}$, the dashed
lines denote their high temperature behaviour in the confining
oscillator ($\hbar\omega$).}{CC-2}
\end{figure} 

The critical temperature $T_c$, which can be estimated from the
disappearance of the coexistence phase in figure~\xref{CC-2}, is
about $10~\MeV$ and has to be compared to the results of ref.
\cite{JMZ84,BLV84,CaY96} for finite nuclei including
Coulomb and surface effects. All authors report a week
dependence of the critical temperature on the mass number in the
region from calcium to lead. Jaqaman's result with the Skyrme
ZR3 interaction \cite{JMZ84} can be extrapolated to
\element{16}{O} to give $T_c\approx8~\MeV$, Bonche \cite{BLV84}
arrives at the same number using the SKM interaction, but gets
$T_c\approx11~\MeV$ with the SIII interaction. Close to the last
result is the value extrapolated from ref. \cite{CaY96} where
$T_c\approx11.5~\MeV$ for Gogny's D1 interaction.

%
%

\begin{samepage}
\vbox{\vspace{5mm}}
{\bf Acknowledgments}\\[5mm]
The authors would like to thank the INT at the University of
Washington/Seattle for the warm hospitality and G.~Bertsch for
stimulating discussions.  This work was supported by a grant of
the CUSANUSWERK to J.~S..
\end{samepage}

\end{document}